\documentclass[aps,pra,notitlepage,10pt,nofootinbib,floatfix,noeprint,twocolumn]{revtex4-2}
\pdfoutput=1
\usepackage[utf8]{inputenc}
\usepackage[english]{babel}
\usepackage[T1]{fontenc}
\usepackage{mplmacros}
\usepackage{thmtools}
\renewcommand{\paragraph}[1]{\addcontentsline{toc}{section}{#1}\emph{#1.}---}

\begin{document}
\title{Randomized routing: ruling out classical networks without trusting source independence}

\author{Martin Pl\'{a}vala}
\email{martin.plavala@itp.uni-hannover.de}
\affiliation{Institut für Theoretische Physik, Leibniz Universität Hannover, 30167 Hannover, Germany}

\begin{abstract}
Network nonlocality without inputs assumes that the sources are independent, but no data can test this, since one classical source feeding all parties reproduces any correlation. Here we propose a randomized routing strategy: a source is duplicated, and the two parties it connects choose at random which copy to read. Whether any classical network, even with correlated sources, explains the routed data is then a linear program, and its infeasibility certificates are linear inequalities on the data. Source independence is thus traded for locality and free choice of the routing, which experiments can enforce. In the triangle scenario, the RGB4 distribution fails the test over a wide range of parameters, including a point where its triangle data alone have a numerically found model with independent classical sources, and it still fails with 3\% white noise at each source.
\end{abstract}

\maketitle

\paragraph{Introduction}
Picture three laboratories at the corners of a triangle, each pair connected by its own source of entangled particles. Every laboratory performs the same measurement in every round: nobody chooses a setting, nobody turns a knob. You might expect that there is then nothing nonclassical to see, since a Bell test needs settings. But if the three sources are independent, some of the resulting correlations $p(abc)$ cannot be reproduced by three independent classical sources wired in the same triangle~\cite{fritz2012beyond,renou2019genuine,tavakoli2022bell}. This is called network nonlocality without inputs, and it has already been put to work: it certifies randomness~\cite{sekatski2023partial}, and it has been proposed for device-independent key distribution~\cite{davino2026device}.

\begin{figure}
\centering
\includegraphics[width=0.66666\columnwidth]{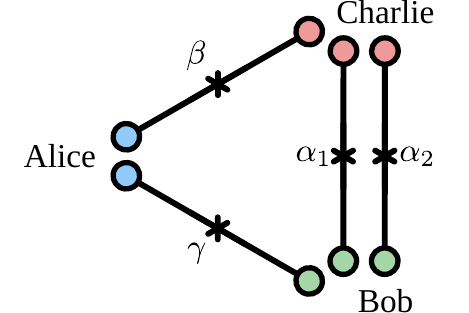}
\caption{Randomized routing in the triangle network. Stars are sources and circles are the systems received by the parties. The source $\alpha$ between Bob ($B$) and Charlie ($C$) is duplicated into two physical copies, $\alpha_1$ and $\alpha_2$. In every round $B$ and $C$ each choose at random which copy to read, and measure its system together with the system from their other source, while the system from the other copy is discarded. If both read the same copy, the round reproduces the triangle data; if they read different copies, it gives new data.}
\label{fig:schematic}
\end{figure}

But here is the catch. Suppose that, hidden behind the triangle, a single classical source samples $(a, b, c)$ from $p$ and sends each laboratory its outcome. It reproduces any $p(abc)$ whatsoever: one hidden source can fake any network. Without inputs, no amount of data tells the two apart, so source independence stays an assumption. And it is not an idle one, since sources may share a laser~\cite{polino2023experimental}, a clock~\cite{supic2020quantum}, or a power supply~\cite{branciard2010characterizing}. As one experimental team put it, ``justifying the independence of different sources turns out to be very demanding''~\cite{polino2023experimental}.

The loophole has not gone unnoticed. It was shown in~\cite{supic2020quantum} that network nonlocality survives when the sources share a common classical variable, provided that the dependence it creates is bounded. For suitable noiseless distributions an arbitrarily small amount of independence then suffices. But with noise a finite amount must be assumed~\cite{meskine2025experimental}, and it cannot be checked: the bounds on the dependence are not observable and can only be estimated under further assumptions, as noted in the experiment implementing this test~\cite{meskine2025experimental}. Another line of work certifies network nonlocality while trusting the network structure only around a few neighbouring parties~\cite{boreiri2025topologically}, and still assumes independent sources there. In all these approaches, some independence is trusted, not tested. Experiments have also made source independence plausible by separating independently pumped sources space-like~\cite{sun2019experimental,gu2023experimental}, which excludes influences between the sources, but not a common cause in their shared past.

So what do you do with an assumption that no data can test? You think outside the box, or rather outside the network. We propose a randomized routing strategy: build one spare copy of a source, and let the two parties it connects choose at random, in every round, which copy to read (Fig.~\ref{fig:schematic}). One hidden source can fake any network, but it cannot always fake one that is rewired at random. It has to fix in advance what every party would answer for every routing choice, since it does not know which copies will be read. Rounds in which both parties read the same copy reproduce the triangle data, while rounds in which they read different copies give new data. A single classical source must explain both with one joint distribution of all counterfactual outcomes, and whether such a distribution exists is a linear program.

In this paper we show that this one test rules out every classical network at once, whether its sources are independent or correlated, through noise-robust linear inequalities on the routed data. Source independence is thereby traded for locality and free choice of the routing, which can be enforced by space-like separation, as in the loophole-free Bell tests~\cite{hensen2015loophole,giustina2015significant,shalm2015strong}.
In the triangle with one spare source, the routed data of the Fritz and elegant joint measurement distributions still admit a single classical source. For the RGB4 distribution we rule out every classical explanation over a wide range of parameters, even at a point where its triangle data alone have a numerically found model with independent classical sources, and still with 3\% white noise at each source.

\paragraph{Randomized routing}
Consider the triangle network with parties $A$, $B$, and $C$, in which source $\alpha$ is shared by $B$ and $C$, $\beta$ by $C$ and $A$, and $\gamma$ by $A$ and $B$. Each party performs one fixed measurement on the two systems it receives and records an outcome $a$, $b$, or $c$. The randomized routing strategy adds a spare copy of one source, say $\alpha$, so that there are two copies $\alpha_1$ and $\alpha_2$, see Fig.~\ref{fig:schematic}. In every round, $B$ and $C$ each choose at random which copy to read, $j, k \in \{1, 2\}$, and feed the system from that copy into their unchanged measurement, while the system from the other copy is discarded inside the party's laboratory. This gives the routed data $p(abc|jk)$. The spare is built as a second, identical source: both copies emit the same state, each its own pairs. Then closed rounds, $j = k$, reproduce the triangle data, $p_{\mathrm{closed}}(abc) = p(abc|jj) = p(abc)$. Open rounds, $j \neq k$, give new data $p_{\mathrm{open}}(abc) = p(abc|jk)$, since $B$ and $C$ then hold halves of two different entangled pairs. The test below takes the routed data as measured, whether or not the copies are identical.

The spare source is real hardware, one more box on the table, not a fictitious copy as in inflation~\cite{wolfe2019inflation}.
Besides it, each of the two parties needs a switch, that is, a device that selects which source they are connected to. Fibre switches driven by random signals, with switching times of about $8 \unit{ns}$, have already been used inside a triangle experiment~\cite{polino2023experimental},
and a research prototype of an optical switch for quantum networks, announced by Cisco, reconfigures in as little as $1 \unit{ns}$~\cite{cisco2026universal}. The routing bits can come from a quantum random number generator. Switching between fibres is thus possible to implement, and every party keeps its single measurement. In the future, dense networks offer several paths between any two nodes~\cite{mariani2025quantum}; we expect quantum networks that connect several places to have more than one source between some parties, and switches to route between them, so the spare source and the switches will likely be part of the quantum internet infrastructure~\cite{kimble2008quantum,wehner2018quantum}.

Nothing in the strategy depends on duplicating only one source. At level $(n_\alpha, n_\beta, n_\gamma)$ each source has that many copies, and each party chooses one copy of each of its two sources: $A$ chooses $i = (i_\beta, i_\gamma)$, $B$ chooses $j = (j_\gamma, j_\alpha)$, and $C$ chooses $k = (k_\alpha, k_\beta)$, with $i_\beta \in \{1, \ldots, n_\beta\}$ and so on. The routed data are $p(abc|ijk)$. A routing choice is closed at a source if both parties of that source read the same copy of it, and open otherwise. The set-up above is level $(2, 1, 1)$, where $j$ and $k$ reduce to $j_\alpha$ and $k_\alpha$. We need this generality for the statements below; all examples use one spare source.

The hypothesis we want to rule out is the one that fakes any network: one classical source sends the same variable $\lambda$ to all three parties, and each party may respond to its own routing choice, but not to the choices of the others, which are made independently of $\lambda$.

\begin{definition}[Single classical source]\label{def:single-source}
Routed data $p(abc|ijk)$ at level $(n_\alpha, n_\beta, n_\gamma)$ admit a single classical source if there are a probability measure $\mu$ on a measurable space $\Lambda$ and conditional distributions $p(a|i,\lambda)$, $p(b|j,\lambda)$, and $p(c|k,\lambda)$, measurable in $\lambda$, such that
\begin{equation}\label{eq:single-source}
p(abc|ijk) = \int_\Lambda p(a|i,\lambda) p(b|j,\lambda) p(c|k,\lambda) \mu(\dd \lambda)
\end{equation}
for all $i$, $j$, $k$.
\end{definition}

The measure $\mu$ does not depend on the routing choices (free choice), and each response depends on the routing choice of its own party only (locality). A party may know which copy it reads: the system it receives could carry the copy label, so its response is allowed to depend on it. Whether a single classical source exists is decided by a linear program. Let $\vec a = (a_i)_i$ collect one outcome of $A$ for each of its routing choices, and similarly $\vec b$ and $\vec c$.

\begin{restatable}{proposition}{propone}\label{prop:lp}
Routed data admit a single classical source if and only if there is a probability distribution $q(\vec a\, \vec b\, \vec c)$ whose marginal on $(a_i, b_j, c_k)$ equals $p(abc|ijk)$ for all $i$, $j$, $k$. If no such $q$ exists, there is a linear inequality on the routed data that every single classical source satisfies and $p$ violates.
\end{restatable}

The entries of $q$ are counterfactual: in every round each party reads one copy only, so only one entry of each of $\vec a$, $\vec b$, and $\vec c$ is ever observed. Proposition~\ref{prop:lp} is the routed analogue of Fine's theorem~\cite{fine1982hidden},
and the inequality is the dual certificate of the linear program. The proof is in Appendix~\ref{app:lp}.

Why rule out a single classical source, and not a specific network topology? Because a single source contains every classical network: grouping all classical sources into one big source gives a special case of Definition~\ref{def:single-source}. Conversely, a single classical source is a classical network whose sources are all collected together. The routed data of a classical network therefore never rule out a single classical source.

\begin{restatable}{proposition}{proptwo}\label{prop:classical-pass}
Let the routed data at level $(n_\alpha, n_\beta, n_\gamma)$ be produced by classical sources: one variable for each copy of each source, with any joint distribution, independent or correlated. Let each party's outcome depend only on its own routing choice and on the variables of all sources and all their copies, including copies it does not read, whose systems are discarded inside its laboratory, and let the routing choices be jointly independent of all source variables. Then the routed data admit a single classical source, that is, the linear program of Proposition~\ref{prop:lp} is feasible.
\end{restatable}

Read backwards, Proposition~\ref{prop:classical-pass} is the statement we use: if no $q$ exists, then no classical sources, independent or correlated, explain the routed data, given locality and free choice of the routing. 

Do we need both kinds of rounds? Yes: rounds that all give the same data never rule out a single classical source.

\begin{restatable}{proposition}{propthree}\label{prop:one-kind}
Let $S$ be a set of triples $(i, j, k)$ of routing choices of $A$, $B$, and $C$, and let the routed data be the same distribution at every triple in $S$, $p(abc|ijk) = p_0(abc)$ for all $(i, j, k) \in S$. Then these data admit a single classical source, that is, Eq.~\eqref{eq:single-source} holds for all $(i, j, k) \in S$.
\end{restatable}

With identical copies, examples of such $S$ are the rounds closed at every source, which all give the triangle data, and, when only $\alpha$ is duplicated, the rounds open at $\alpha$, which all give $p_{\mathrm{open}}$. The single source draws one sample $(a, b, c)$ of $p_0$ and gives each party the same outcome whatever copies it reads, so the routing has no effect. All the power of randomized routing therefore comes from comparing rounds whose data differ. Both proofs are in Appendix~\ref{app:lp}.

\paragraph{Requirement and assumptions}
Randomized routing has one requirement and rests on three assumptions. The requirement is the spare source: one more physical box, so that there are four physical sources instead of three. It is a requirement, not an independence assumption, since by Proposition~\ref{prop:classical-pass} the spare may be correlated with all the other sources; it only has to exist. Such spares may be in place anyway, as argued above.

First assumption, locality: the outcome of a party does not depend on the routing choices of the others. Second, free choice: the routing choices are jointly independent of the sources. Both can be enforced by space-like separation, generating each routing bit space-like separated from the emission events at all four sources and from the measurements of the other parties, as in the loophole-free Bell tests~\cite{hensen2015loophole,giustina2015significant,shalm2015strong}, with the usual caveat that superdeterminism can never be excluded. Third, independent and identically distributed rounds, as is standard in network nonlocality; memory effects across rounds~\cite{weilenmann2025memory} are beyond our scope. As in any Bell test, the detection loophole must be closed~\cite{hensen2015loophole,giustina2015significant,shalm2015strong}.

One may object that we have only traded one independence assumption for another. Source independence has been argued to be not much different from free choice of the settings~\cite{branciard2010characterizing}, and called the network analogue of measurement independence~\cite{davino2026device}, and free choice is a form of measurement independence. But the two are of different kinds. Free choice concerns a bit that the experimenter generates, randomizes, and can separate from everything else in spacetime. Source independence concerns the hidden internals of devices that the experimenter does not control, and without inputs no data can check it. Randomized routing turns an assumption of the second kind into one of the first.

\paragraph{Bell tests and inflation}
By Proposition~\ref{prop:lp}, the routed data that admit a single classical source form the local polytope of a three-party Bell scenario in which the routing choices are the settings: $A$ has $n_\beta n_\gamma$ of them, $B$ has $n_\gamma n_\alpha$, and $C$ has $n_\alpha n_\beta$. Randomized routing could therefore be seen as equivalent to a Bell test, but one whose settings are implemented by switching between copies of a source rather than by changing measurements, and whose closed rounds are the network data themselves. In the simplest case, level $(2, 1, 1)$, $A$ has no routing choice, and the test can be read as a Bell test between $B$ and $C$, for each outcome $a$ of $A$ separately, as in event-ready schemes~\cite{zukowski1993event,branciard2010characterizing}, with no outcome of $A$ discarded. This bipartite reading breaks down as soon as a second source, between $A$ and $B$ or between $A$ and $C$, is duplicated as well, since $A$ then has routing choices too; the test is then a three-party Bell test, which the formulation above and its linear program handle equally directly.

Neither is randomized routing inflation~\cite{wolfe2019inflation,navascues2020inflation}, a technique that tests network models by imagining several copies of each source and asking for a consistent joint distribution of the resulting outcomes. The unknown $q$ is the same as in a fanout inflation with the same number of copies, one in which a single copy of a source feeds several copies of the same party, but the constraints differ. Inflation fixes the marginals on sub-networks shaped like the triangle to $p$ and imposes factorisation on copies that share no source, which needs independent sources, while the open three-party marginals stay free. Randomized routing measures the open marginals and imposes no factorisation. Linear programs are already a standard tool for network nonlocality~\cite{wolfe2019inflation,pozas2023proofs,hayes2025linear}, and the two tests can run on the same data: inflation on the closed rounds tests network nonlocality, and randomized routing on all rounds rules out every classical explanation.

\paragraph{Examples}
We test three distributions from the literature, always at level $(2, 1, 1)$, with one spare copy of $\alpha$. This is the simplest extension of the triangle: levels such as $(2, 2, 1)$ or $(3, 1, 1)$ need a second spare source, which increases the requirements on the experiment. In each case we compute the closed and open data from the states and measurements that produce them. For the Fritz distribution~\cite{fritz2012beyond}, a CHSH test on one entangled source embedded in the triangle, with the other two sources carrying the setting bits, the open data have a simple structure: $b$ and $c$ are independent given $a$, $p_{\mathrm{open}}(abc) = p(a) p(b|a) p(c|a)$. Here $p(b|a)$ and $p(c|a)$ are the conditionals of the closed data, equal to those of the open data by locality. Then
\begin{equation}\label{eq:fritz-q}
q(a\, b_1 b_2\, c_1 c_2) = p(a) p_{\mathrm{closed}}(b_1 c_1|a) p_{\mathrm{closed}}(b_2 c_2|a)
\end{equation}
reproduces all routed data: the single source draws $a$, and then one closed pair $(b, c)$ for each copy, independently. The same construction works at every level $(n, 1, 1)$, and, with the roles of the parties exchanged, whichever source is duplicated.

For the elegant joint measurement~\cite{gisin2017elegant,gisin2019entanglement}, every source emits a singlet and every party performs a four-outcome entangled measurement on its two qubits. Its triangle data were recently shown to admit no model with independent classical sources~\cite{gitton2025elegant}, but its routed data admit a single classical source. Appendix~\ref{app:ejm} gives an exact $q$.

The RGB4 family~\cite{renou2019genuine,pozas2023proofs}
behaves differently. In it every source emits $\ket{\psi^+}$, and every party measures its two qubits in the basis $\ket{00}$, $u \ket{01} + \sqrt{1 - u^2} \ket{10}$, $\sqrt{1 - u^2} \ket{01} - u \ket{10}$, $\ket{11}$, with four outcomes. Here all verdicts and range endpoints are numerical. The routed data admit no single classical source for $u^2$ in about $(0.08, 0.92)$, except at $u^2 = \frac{1}{2}$, where the routed data admit a single classical source. The range is symmetric under $u^2 \leftrightarrow 1 - u^2$, which maps the family to itself up to exchanging $B$ and $C$. Compare this with the triangle data alone. Their network nonlocality, which assumes independent sources, is proven for $u^2 \in (0.785, 1)$~\cite{renou2019genuine} and $u^2 \in [0.5631, 0.6563]$~\cite{pozas2023proofs}, and, by the same symmetry, for their mirror images. At $u^2 = \frac{1}{2}$ they have an explicit model with independent classical sources, and at the endpoint $u^2 \approx 0.785$ a numerical one~\cite{renou2019genuine}. So at $u^2 \approx 0.785$ the triangle data alone have a numerically found model with independent classical sources, and still no classical network, independent or correlated, explains the routed data. Conversely, for $u^2 \in (0.92, 1)$ the triangle data are network nonlocal, while the routed data admit a single classical source. The two tests answer different questions, under different assumptions, and routing uses more data.

\begin{figure}
\centering
\includegraphics[width=\columnwidth]{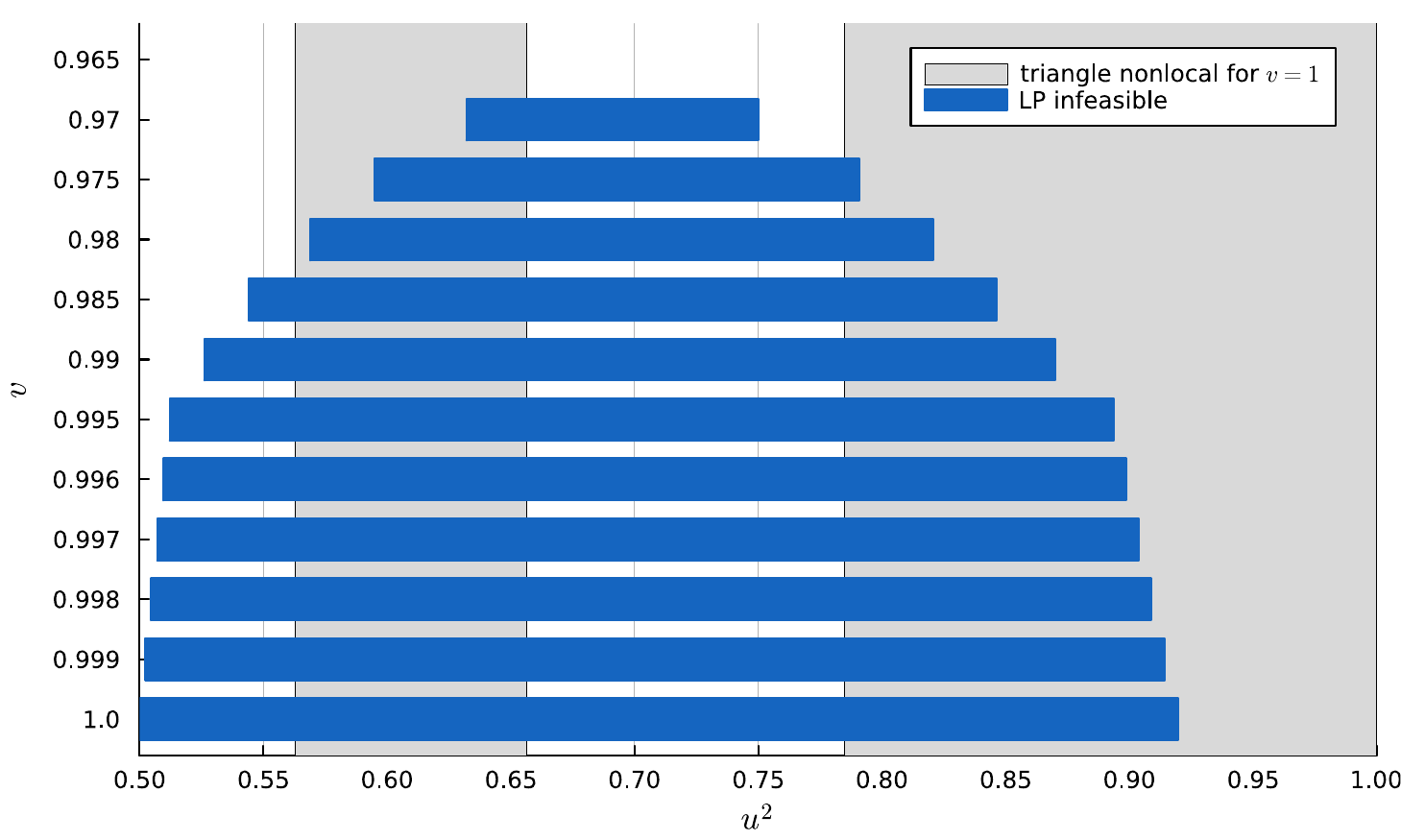}
\caption{Routed data of the RGB4 family with one spare source, for white noise of visibility $v$ at each of the four sources, $u^2 \in [\frac{1}{2}, 1]$; the range $[0, \frac{1}{2}]$ is its mirror image. Coloured: values of $u^2$ at which a single classical source is ruled out by a checked dual certificate; elsewhere a feasible point was found, apart from a few unresolved points. Grey bands: proven network nonlocality of the noiseless triangle data alone, for $v = 1$ only~\cite{renou2019genuine,pozas2023proofs}.}
\label{fig:noise}
\end{figure}

Fig.~\ref{fig:noise} shows how noise shrinks the range. With white noise of visibility $v$, each of the four sources emits $v \ketbra{\psi^+} + (1 - v) \frac{\I}{4}$. A single classical source is still ruled out for some $u^2$ at $v = 0.97$ ($3\%$ white noise), and nowhere at $v = 0.965$. The most robust point, $u^2 \approx 0.69$, lies where the network nonlocality of the triangle data is not known. For comparison, proofs of network nonlocality for distributions of this kind tolerate only small amounts of white noise~\cite{boreiri2025noise}, and numerics place the threshold for the RGB4 triangle data near $v = 0.99$ at $u^2 = 0.85$~\cite{girardin2026versatile}. Triangle experiments report source fidelities up to about $0.98$~\cite{wang2026experimental,meskine2025experimental}, that is, visibilities up to about $0.97$, since $F = \frac{1 + 3 v}{4}$ for white noise. Current sources are thus at the threshold before any measurement or switching noise, which reduce the margin further~\cite{wang2026experimental}.

All programs were solved with Mosek. Every infeasibility rests on a dual certificate checked outside the solver, and every feasible verdict on a point satisfying the constraints to $10^{-5}$, see Appendix~\ref{app:numerics} for more details.

\paragraph{Many copies}
Infeasibility at one level persists at every higher level when the copies are identical, since the constraints of the lower level are among those of the higher one. What does the hierarchy of levels $(n, 1, 1)$ rule out? For one duplicated source the answer is simple. At level $(n, 1, 1)$ the source $\alpha$ has $n$ copies, and $B$ and $C$ choose $j, k \in \{1, \ldots, n\}$. With identical copies, $p(abc|jk)$ equals $p_{\mathrm{closed}}(abc)$ for $j = k$ and $p_{\mathrm{open}}(abc)$ for $j \neq k$.

\begin{restatable}{proposition}{propfour}\label{prop:definetti}
The data $p_{\mathrm{closed}}$ and $p_{\mathrm{open}}$ admit a single classical source at every level $(n, 1, 1)$ if and only if, for every outcome $a$, there are a probability measure $\mu_a$ on a measurable space $\Lambda$ and conditional distributions $r(bc|a,\lambda)$, measurable in $\lambda$, such that
\begin{align}
p_{\mathrm{closed}}(abc) &= p(a) \int_\Lambda r(bc|a,\lambda) \mu_a(\dd \lambda), \label{eq:definetti-closed} \\
p_{\mathrm{open}}(abc) &= p(a) \int_\Lambda r(b|a,\lambda) r(c|a,\lambda) \mu_a(\dd \lambda), \label{eq:definetti-open}
\end{align}
where $p(a)$ is the marginal of $A$, the same in closed and open data by locality, and $r(b|a,\lambda)$ and $r(c|a,\lambda)$ are the marginals of $r(bc|a,\lambda)$.
\end{restatable}

Since $A$ never routes, its outcome can be taken as part of the source: in such a model a global classical variable, $a$ together with $\lambda$, reaches all three parties, and given it every copy of $\alpha$ acts as an independent classical source emitting $(b, c)$ with distribution $r(bc|a,\lambda)$. If the routed data have no such model, some finite number of copies rules out a single classical source. The proof, stated in full in Appendix~\ref{app:definetti}, symmetrises a feasible $q$ over the copies of $\alpha$, uses compactness to obtain a consistent sequence over all levels~\cite{aubrun2025monogamy}, and applies the de Finetti theorem~\cite{hewitt1955symmetric,aubrun2025monogamy}. The converse is a direct construction. Finite de Finetti theorems~\cite{diaconis1980finite} would show how fast the levels converge, with corrections of order $\frac{1}{n}$ for fixed numbers of outcomes, and the theorem of Aldous and Hoover~\cite{aldous1981representations,hoover1979relations} could treat all three sources at once. We leave both to future work.

\paragraph{Discussion}
Source independence need not be trusted: it can be traded for one more box and a random switch. It is a critical, untestable assumption of network nonlocality without inputs, and randomized routing replaces it by assumptions that experiments can enforce, with hardware of the kind quantum networks are building. We have shown that the nonclassicality of the data of a routed network experiment can be established in this way. With one physical spare source and random routing between its copies, a linear program decides whether any classical network, with possibly correlated sources, explains the routed data, given locality and free choice of the routing. Neither the closed nor the open rounds suffice alone, and for one duplicated source the hierarchy over the number of copies has a de Finetti characterisation.

For the RGB4 family, the routed data rule out every classical explanation over a wide range of parameters, even at a point where the triangle data alone have a numerically found model with independent classical sources, and still with 3\% white noise at each source.

The main open question concerns quantum sources. A single quantum source shared by all parties reproduces the routed data of any quantum network, so randomized routing cannot rule it out; which quantum explanations it can exclude remains unknown. Routing at higher levels $(n, 1, 1)$, or with more duplicated sources, might also show that the elegant joint measurement is nonclassical without assuming independent sources. Duplicating more sources is possible, at the cost of more hardware. Finite statistics and memory effects across rounds also remain to be treated. Finally, protocols built on network nonlocality without inputs~\cite{sekatski2023partial,davino2026device}
currently assume independent sources, and routed data might let them drop that assumption.

\paragraph{Acknowledgements}
I used large language models to assist with literature search, numerical and analytical checks, and drafting and editing the text. I take full responsibility for the content of this paper.

I acknowledge support from the Niedersächsisches Ministerium für Wissenschaft und Kultur.

The code and the infeasibility certificates are available upon request.

\bibliography{citations}

\onecolumngrid

\appendix

\section{The linear program and the proofs of Propositions~\ref{prop:lp}, \ref{prop:classical-pass}, and \ref{prop:one-kind}}\label{app:lp}

Fix the level $(n_\alpha, n_\beta, n_\gamma)$ and finite outcome sets. The routing choices are $i = (i_\beta, i_\gamma) \in I_A = \{1, \ldots, n_\beta\} \times \{1, \ldots, n_\gamma\}$ for $A$, $j = (j_\gamma, j_\alpha) \in I_B = \{1, \ldots, n_\gamma\} \times \{1, \ldots, n_\alpha\}$ for $B$, and $k = (k_\alpha, k_\beta) \in I_C = \{1, \ldots, n_\alpha\} \times \{1, \ldots, n_\beta\}$ for $C$. The variables are $q(\vec a\, \vec b\, \vec c) \geq 0$, with $\vec a = (a_i)_{i \in I_A}$, $\vec b = (b_j)_{j \in I_B}$, and $\vec c = (c_k)_{k \in I_C}$. For every $(i, j, k) \in I_A \times I_B \times I_C$ and all outcomes $a$, $b$, $c$ we impose
\begin{equation}\label{eq:lp-constraints}
\sum_{\substack{\vec a, \vec b, \vec c: \\ a_i = a, b_j = b, c_k = c}} q(\vec a\, \vec b\, \vec c) = p(abc|ijk).
\end{equation}
Normalisation of $q$ follows from any one choice $(i, j, k)$. With $d$ outcomes per party the program has $d^{n_\beta n_\gamma + n_\gamma n_\alpha + n_\alpha n_\beta}$ variables and $d^3 (n_\alpha n_\beta n_\gamma)^2$ equality constraints;
 at level $(2, 1, 1)$ with $d = 4$ these are $1024$ and $256$.

If the program is feasible at level $(n_\alpha, n_\beta, n_\gamma)$, it is feasible at every level $(n'_\alpha, n'_\beta, n'_\gamma)$ with $n'_\alpha \leq n_\alpha$, $n'_\beta \leq n_\beta$, and $n'_\gamma \leq n_\gamma$, provided the data at the higher level, at the routing choices within the smaller index sets, are the data of the lower level: keep the entries of $q$ whose copy indices lie in the smaller index sets and sum out the others. The constraints of the lower level are among those of the higher one, with the same data. Equivalently, infeasibility persists at every higher level.

\propone*

\begin{proof}
If the data admit a single classical source, then
\begin{equation}
q(\vec a\, \vec b\, \vec c) = \int_\Lambda \prod_{i \in I_A} p(a_i|i,\lambda) \prod_{j \in I_B} p(b_j|j,\lambda) \prod_{k \in I_C} p(c_k|k,\lambda) \mu(\dd \lambda)
\end{equation}
has the required marginals, since summing over all entries except $a_i$, $b_j$, $c_k$ turns every other factor into $1$. Conversely, given $q$, let the single source distribute a sample $(\vec a\, \vec b\, \vec c)$ of $q$ to all parties, and let each party output the entry selected by its routing choice.

If no $q$ exists, linear programming duality gives coefficients $y_{ijk}(abc)$ with $\sum_{i,j,k,a,b,c} y_{ijk}(abc) p(abc|ijk) < 0$, while $\sum_{i,j,k} y_{ijk}(a_i b_j c_k) \geq 0$ for every tuple $(\vec a\, \vec b\, \vec c)$. Every single classical source is a mixture of deterministic single sources, each outputting a fixed tuple $(\vec a\, \vec b\, \vec c)$, so it satisfies the inequality.
\end{proof}

\proptwo*

\begin{proof}
Let $\alpha_m$, $\beta_m$, $\gamma_m$ also denote the classical variables emitted by the copies, $\nu$ their joint distribution, which is arbitrary, and let the responses be $p(a|i, \vec\alpha, \vec\beta, \vec\gamma)$, $p(b|j, \vec\alpha, \vec\beta, \vec\gamma)$, and $p(c|k, \vec\alpha, \vec\beta, \vec\gamma)$. Since the routing choices are jointly independent of the source variables,
\begin{equation}
p(abc|ijk) = \int p(a|i, \vec\alpha, \vec\beta, \vec\gamma) p(b|j, \vec\alpha, \vec\beta, \vec\gamma) p(c|k, \vec\alpha, \vec\beta, \vec\gamma) \nu(\dd \vec\alpha \dd \vec\beta \dd \vec\gamma).
\end{equation}
This is Definition~\ref{def:single-source} with $\lambda = (\vec\alpha, \vec\beta, \vec\gamma)$, $\mu = \nu$, and $p(a|i,\lambda) = p(a|i, \vec\alpha, \vec\beta, \vec\gamma)$, and similarly for $B$ and $C$. By Proposition~\ref{prop:lp} the program is feasible.
\end{proof}

This covers in particular a classical triangle network with independent sources, run with independent, identical copies and responses that depend only on the copies read and ignore the routing choice. The data at a routing choice then depend only on the sources at which the routing choice is closed. If such a network reproduces the triangle data and the data for every set of closed sources that occurs at the level, it reproduces the routed data, which therefore admit a single classical source.

\propthree*

\begin{proof}
Let $p(abc|ijk) = p_0(abc)$ for all $(i, j, k) \in S$. Set $q(\vec a\, \vec b\, \vec c) = p_0(abc)$ if all entries of $\vec a$ equal $a$, all entries of $\vec b$ equal $b$, and all entries of $\vec c$ equal $c$, and $q = 0$ otherwise. Its marginal on $(a_i, b_j, c_k)$ is $p_0$ for every routing choice $(i, j, k)$, so this construction reproduces $p_0$ at every routing choice, in particular at those in $S$. By Proposition~\ref{prop:lp}, applied to the data $p_0$ at every routing choice, these data admit a single classical source.
\end{proof}

\section{The elegant joint measurement}\label{app:ejm}

We show that the routed data of the elegant joint measurement at level $(2, 1, 1)$ admit a single classical source. All three sources emit singlets, and each party performs the elegant joint measurement of~\cite{gisin2017elegant} on its two qubits, with the same qubit ordering as in Appendix~\ref{app:numerics}; see also~\cite{gisin2019entanglement,tavakoli2022bell}. The measurement has four outcomes. The closed data are the triangle data,
\begin{equation}
p_{\mathrm{closed}}(abc) =
\begin{cases}
\frac{25}{256} & a = b = c, \\
\frac{1}{256} & \text{exactly two outcomes equal}, \\
\frac{5}{256} & a, b, c \text{ all different}.
\end{cases}
\end{equation}
In the open rounds $B$ and $C$ read different copies of $\alpha$, and
\begin{equation}
p_{\mathrm{open}}(abc) =
\begin{cases}
\frac{13}{256} & a = b = c, \\
\frac{5}{256} & a = b \neq c \text{ or } a = c \neq b, \\
\frac{1}{256} & b = c \neq a, \\
\frac{3}{256} & a, b, c \text{ all different}.
\end{cases}
\end{equation}
Let $R(abc) = \min(p_{\mathrm{closed}}(abc), p_{\mathrm{open}}(abc))$, so $R(abc)$ is $\frac{13}{256}$, $\frac{1}{256}$, and $\frac{3}{256}$ in the three cases of $p_{\mathrm{closed}}$. Let $\varepsilon_{abc} = 1$ if $a$, $b$, $c$ are all different and $\varepsilon_{abc} = 0$ otherwise, and let $b_1, b_2$ and $c_1, c_2$ be the outcomes of $B$ and $C$ on the two copies. Then
\begin{equation}\label{eq:ejm-q}
q(a\, b_1 b_2\, c_1 c_2) = \delta_{b_1 b_2} \delta_{c_1 c_2} R(a b_1 c_1) + \frac{2}{256} \left( \delta_{a b_1} \delta_{a c_1} \varepsilon_{a b_2 c_2} + \delta_{a b_2} \delta_{a c_2} \varepsilon_{a b_1 c_1} \right)
\end{equation}
reproduces all routed data. On the closed marginal $(a, b_1, c_1)$ the first term gives $R(a b_1 c_1)$. The second term adds $\frac{12}{256}$ when $a = b_1 = c_1$, from the six pairs $(b_2, c_2)$ that are different from each other and from $a$, and $\frac{2}{256}$ when $a, b_1, c_1$ are all different, from $b_2 = c_2 = a$. This gives $\frac{25}{256}$, $\frac{1}{256}$, and $\frac{5}{256}$. On the open marginal $(a, b_1, c_2)$ the first term gives $R(a b_1 c_2)$, and the second term adds $\frac{4}{256}$ when $a = b_1 \neq c_2$ and when $a = c_2 \neq b_1$. This gives $\frac{13}{256}$, $\frac{5}{256}$, $\frac{5}{256}$, $\frac{1}{256}$, and $\frac{3}{256}$. The other two marginals follow by exchanging the copies.

\section{Numerics}\label{app:numerics}

The routed data of the RGB4 family are computed from its states and measurements~\cite{renou2019genuine,pozas2023proofs}.
Every source, both copies of $\alpha$ included, emits $v \ketbra{\psi^+} + (1 - v) \frac{\I}{4}$, and every party measures in the basis $\ket{00}$, $u \ket{01} + \sqrt{1 - u^2} \ket{10}$, $\sqrt{1 - u^2} \ket{01} - u \ket{10}$, $\ket{11}$, with $u = \cos t$. Each party orders its two qubits as (source shared with the previous party, source shared with the next) along $A \to B \to C \to A$, that is, $A$ holds $(\beta, \gamma)$, $B$ holds $(\gamma, \alpha)$, and $C$ holds $(\alpha, \beta)$. In open rounds $B$ and $C$ hold halves of different copies of $\alpha$. We scan $t$ in steps of $0.001$, for $v \in \{1, 0.999, 0.998, 0.997, 0.996, 0.995, 0.99, 0.985, 0.98, 0.975, 0.97, 0.965\}$. Figure~\ref{fig:noise} shows $u^2 \in [\frac{1}{2}, 1]$; the range below $\frac{1}{2}$ is its mirror image under the symmetry of the family.

Each linear program, the constraints of Appendix~\ref{app:lp} together with normalisation (normalisation is redundant, see Appendix~\ref{app:lp}, but used in the certificate below), $M q = p$ and $q \geq 0$, is solved with MOSEK 11.2.2 through its Julia interface~\cite{mosek2026julia}, with its interior-point method. We index the variables of the program by $x$ and its constraints by $e$. The verdict is taken outside the solver. A point is infeasible if the dual vector $y$ returned by the solver is a certificate, $y^\top p > \max_x (M^\top y)_x$, with a gap above $10^{-9}$ after normalising $y$ to $\max_e \abs{y_e} = 1$; this is the certificate of Appendix~\ref{app:lp}, up to sign and a multiple of the normalisation constraint. This proves infeasibility: any feasible $q$ satisfies $q \geq 0$ and $\sum_x q_x = 1$, so $y^\top p = y^\top M q \leq \max_x (M^\top y)_x$; floating-point errors are of order $10^{-15}$, far below the threshold. A point is feasible if the primal point returned by the solver satisfies $M q = p$ and $q \geq 0$ up to $10^{-5}$. We test for a certificate first, since some data entries at $v = 0.999$ are as small as $\frac{(1 - v)^3}{64} \approx 1.6 \times 10^{-11}$, far below this tolerance. The infeasible ranges therefore rest on checked certificates, and the feasible verdicts on points that satisfy the constraints to $10^{-5}$.

A few points remain unresolved: there neither a certificate nor a feasible point passes the checks above, which we attribute to numerical artefacts at data entries of the order of the solver tolerance. They are omitted from Fig.~\ref{fig:noise}.

\section{The proof of Proposition~\ref{prop:definetti}}\label{app:definetti}

\propfour*

\begin{proof}
At level $(n, 1, 1)$ we write $q^{(n)}(a\, \vec b\, \vec c)$ with $\vec b = (b_1, \ldots, b_n)$ and $\vec c = (c_1, \ldots, c_n)$, where $b_m$ and $c_m$ are the outcomes of $B$ and $C$ on copy $m$. The constraints of Appendix~\ref{app:lp} require the marginal on $(a, b_j, c_k)$ to be $p_{\mathrm{closed}}$ for $j = k$ and $p_{\mathrm{open}}$ for $j \neq k$.

Assume first that such $\mu_a$ and $r$ exist. Then
\begin{equation}
q^{(n)}(a\, \vec b\, \vec c) = p(a) \int_\Lambda \prod_{m=1}^{n} r(b_m c_m|a,\lambda) \mu_a(\dd \lambda)
\end{equation}
has marginal $p(a) \int_\Lambda r(bc|a,\lambda) \mu_a(\dd \lambda) = p_{\mathrm{closed}}(abc)$ on $(a, b_j, c_j)$, and $p(a) \int_\Lambda r(b|a,\lambda) r(c|a,\lambda) \mu_a(\dd \lambda) = p_{\mathrm{open}}(abc)$ on $(a, b_j, c_k)$ with $j \neq k$. By Proposition~\ref{prop:lp} the data admit a single classical source at every level.

Conversely, assume that the program is feasible at every level. A permutation $\pi$ of $\{1, \ldots, n\}$ acting on the copies maps $(b_m, c_m)$ to $(b_{\pi(m)}, c_{\pi(m)})$. It maps closed constraints to closed ones and open constraints to open ones, so the average of a feasible $q^{(n)}$ over all permutations is feasible and symmetric. The set of symmetric feasible points at level $n$ is compact, and summing out the copies $m > n$ maps symmetric feasible points at a higher level to symmetric feasible points at level $n$ (Appendix~\ref{app:lp}). For every $N$ pick a symmetric feasible point $q^{(N)}$ at level $N$. By compactness and diagonal extraction there is a subsequence along which its marginals on the first $n$ copies converge for every $n$, as in~\cite{aubrun2025monogamy}.
The limits $Q^{(n)}$ are symmetric, feasible, and consistent: summing $Q^{(n+1)}$ over $(b_{n+1}, c_{n+1})$ gives $Q^{(n)}$. Every $Q^{(n)}$ has marginal $p(a)$ on $a$.

By the Kolmogorov extension theorem, the family $Q^{(n)}$ is the distribution of $a$ together with an infinite sequence of pairs $(b_m, c_m)$, and conditioned on any $a$ with $p(a) > 0$ this sequence is exchangeable. The de Finetti theorem~\cite{hewitt1955symmetric,aubrun2025monogamy} gives a probability measure $\mu_a$ on $\Lambda$, here the set of distributions $\lambda$ on pairs $(b, c)$, such that
\begin{equation}
Q^{(n)}(a\, \vec b\, \vec c) = p(a) \int_\Lambda \prod_{m=1}^{n} \lambda(b_m c_m) \mu_a(\dd \lambda).
\end{equation}
Put $r(bc|a,\lambda) = \lambda(bc)$, and choose $\mu_a$ arbitrarily when $p(a) = 0$. Since $Q^{(2)}$ is feasible, its marginal on $(a, b_1, c_1)$ is $p_{\mathrm{closed}}$ and its marginal on $(a, b_1, c_2)$ is $p_{\mathrm{open}}$, which are Eqs.~\eqref{eq:definetti-closed} and~\eqref{eq:definetti-open}.
\end{proof}

\end{document}